\documentclass[numbers]{article}
\usepackage{zeldovich}

\begin{document}

\title{Induced Gravitational Collapse in the BATSE era: the case of GRB 970828}
\author{
R. Ruffini$^{1,2,3,5}$\\
 Jorge A. Rueda$^{1,2}$, C. Barbarino$^{1}$, C. L. Bianco$^{1,2}$\\ 
, H. Dereli$^{3,4}$,M. Enderli$^{1,3}$, L. Izzo$^{1,2}$,\\
 M. Muccino$^{1}$, A. V. Penacchioni$^{5}$, G. B. Pisani$^{1,3}$,  Y. Wang$^{1}$\\
\tiny\itshape
$^{1}$Dip. di Fisica and ICRA, Sapienza Universit\`a di Roma, Piazzale Aldo Moro 5, I-00185 Rome, Italy.\\
\tiny\itshape 
$^{2}$ICRANet, Piazza della Repubblica 10, I-65122 Pescara, Italy.\\
\tiny\itshape 
$^{3}$Universite de Nice Sophia Antipolis, CEDEX 2, Grand Chateau Parc Valrose, Nice, France.\\
\tiny\itshape 
$^{4}$Observatoire de la C\^ote d'Azur, F-06304 Nice, France.\\
\tiny\itshape 
$^{5}$ ICRANet-Rio, Rua Dr. Xavier Sigaud 150, Rio de Janeiro, RJ, 22290-180, Brazil.\\}
\maketitle

\begin{abstract}{Following the recently established ``Binary-driven HyperNova'' (BdHN) paradigm, we here interpret GRB 970828 in terms of the four episodes typical of such a model. The ``Episode 1'', up to 40 s after the trigger time t$_0$, with a time varying thermal emission and a total energy of $E_{iso,1st} = 2.60 \times 10^{53}$ erg, is interpreted as due to the onset of an hyper-critical accretion process onto a companion neutron star, triggered by the companion star, an FeCO core approaching a SN explosion. The ``Episode 2'', observed up t$_0$+90 s, is interpreted as a canonical gamma ray burst, with an energy of $E^{e^+e^-}_{tot} = 1.60 \times 10^{53}$ erg, a baryon load of $B = 7 \times 10^{-3}$ and a bulk Lorentz factor at transparency of $\Gamma = 142.5$.  From this Episode 2, we infer that the GRB exploded in an environment with a large average particle density $\langle n \rangle$  $\approx 10^3$ particles/cm$^3$ and dense clouds characterized by typical dimensions of (4 $\div$ 8) $\times 10^{14}$ cm and $\delta n / n \sim 10$. The ``Episode 3'' is identified from t$_0$+90 s all the way up to 10$^{5-6}$ s: despite the paucity of the early X-ray data, typical in the BATSE, pre-Swift era, we find extremely significant data points in the late X-ray afterglow emission of GRB 970828, which corresponds to the ones observed in all BdHNe sources. The ``Episode 4'', related to the Supernova emission, does not appear to be observable in this source, due to the presence of darkening from the large density of the GRB environment, also inferred from the analysis of the Episode 2.}\end{abstract}

\section{Introduction}\label{sec:1}

The Gamma Ray Burst (GRB) 970828 is one of the first GRBs with an observed X-ray and radio afterglow and a determined redshift of $z$=0.9578 from the identification of its host galaxy \cite{Djorgovski2001}. It was detected by the All Sky Monitor (ASM) detector on board the Rossi X-ray Timing Explorer (RXTE) spacecraft \cite{IAUC6726}, and then observed also by the Burst And Transient Source Experiment (BATSE) on board the Compton Gamma-Ray Observatory \cite{IAUC6728}. The crucial data on the afterglow of GRB 970828 were collected by the Advanced Satellite for Cosmology and Astrophysics (ASCA) in the (2 - 10) keV energy range, one day after the RXTE detection \cite{IAUC6729}, and by ROSAT \cite{IAUC6757} in the (0.1 - 2.4) keV, one week later. Observations on optical wavelengths failed to detect the optical afterglow \cite{IAUC6735,Groot1998}. The fluence measured by BATSE implies an isotropic energy for the total emission of $E_{iso}=4.2 \times 10^{53}$ erg. This source is still presenting today, after 15 years from its discovery, an extremely rich problematic in the identification of its astrophysical nature.

\begin{figure}
\centering      
\includegraphics[width=0.49\hsize,clip]{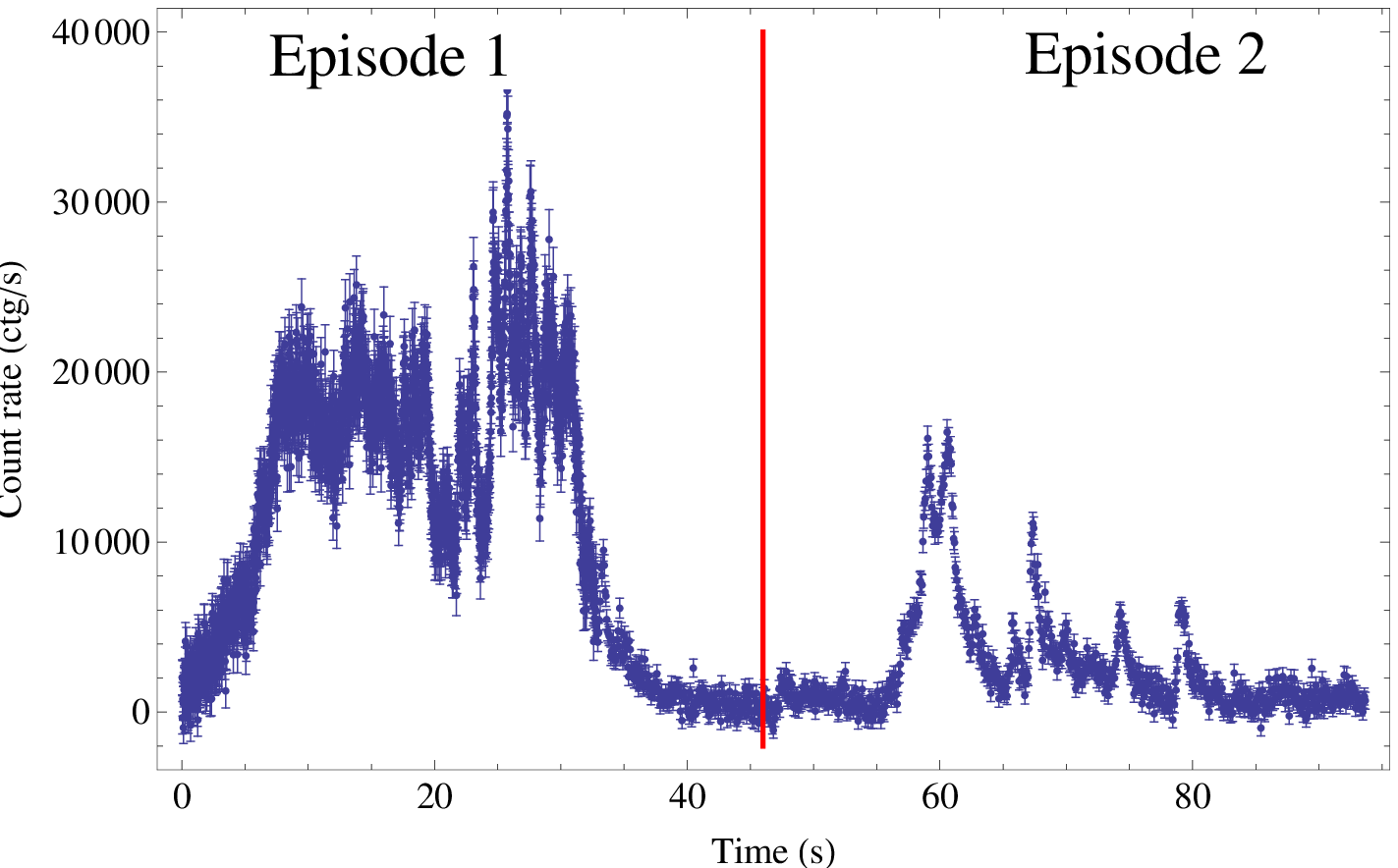}
\includegraphics[width=0.49\hsize,clip]{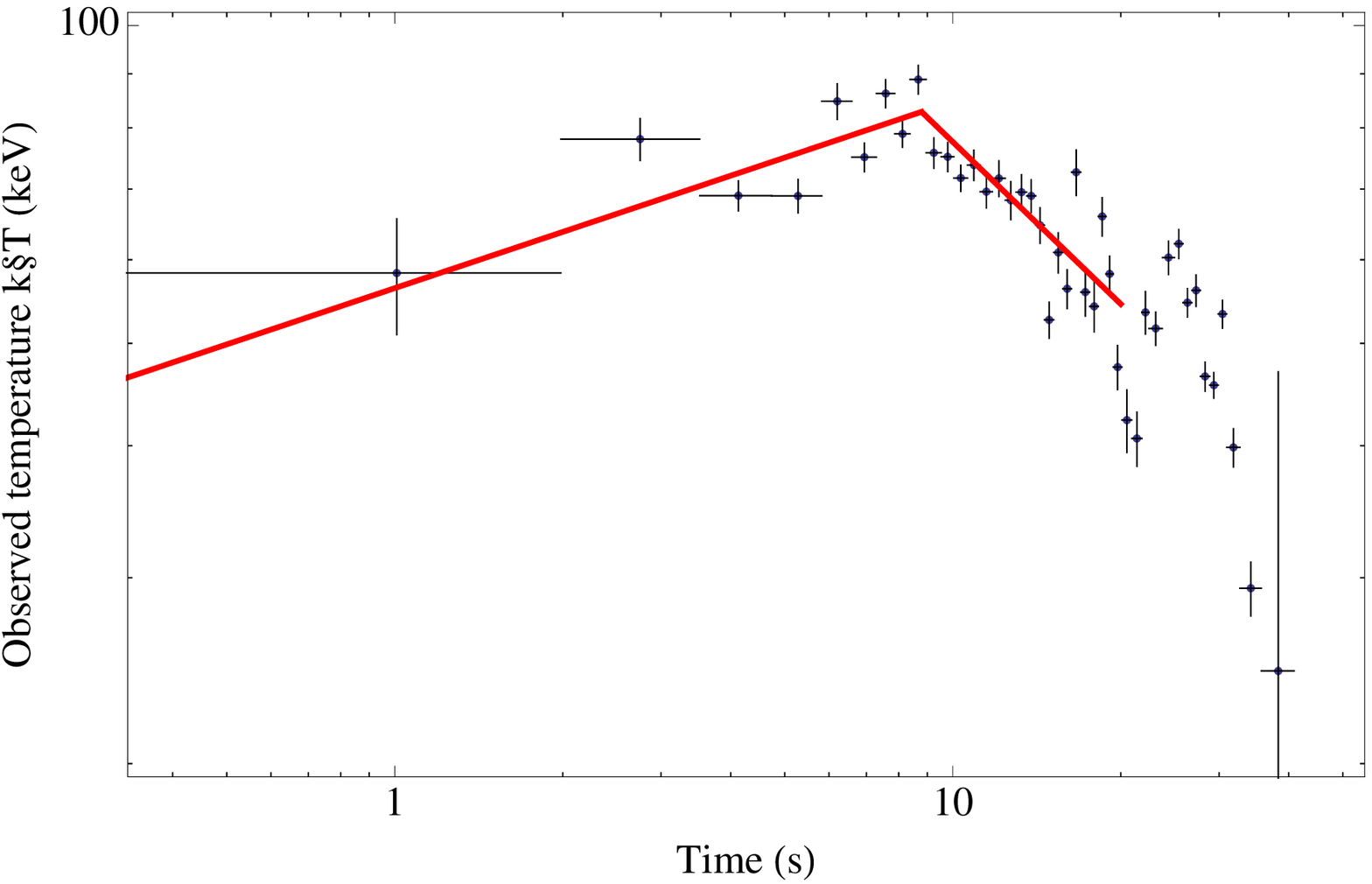}\\
\caption{ \textit{Left panel:} BATSE-LAD light curve of GRB 970828 in the 25-1900 keV energy range. \textit{Right panel:} The behavior of the observed temperature in GRB 970828 and the best fit with a broken power-law function of the first 20 s as presented in \cite{Peer2007}.}
\label{fig:4b.0}
\end{figure}

The recent joint GRB observations made by satellites as Swift \cite{Gehrels2004}, Fermi \cite{Meegan2009}, AGILE \cite{AGILE}, Konus-WIND \cite{KonusWIND} in hard X-rays energy range, as well as the follow-up of their afterglow emission in the (0.3 - 10) keV range by the X-Ray Telescope (XRT) \cite{Burrows2005} on-board Swift, and the corresponding follow-up observations in the optical and radio wavelengths have made possible a new understanding of the entire GRB process. In this paper we start a procedure of revisiting previous GRBs in the BATSE, pre-Swift era, including the new understanding mentioned above.
In particular, we apply to GRB 970828 the new BdHN scenario, in which GRBs associated with Supernovae (SNe) \cite{Ruffini2007b,Rueda2012}, are composed of four different episodes \cite{Izzo2012,Penacchioni2012,Penacchioni2013,Pisani2013}. 
\begin{itemize}
\item The ``Episode 1'' corresponds to the emission from an hyper-critical accretion onto the a neutron star (NS) due to the onset of a Supernova (SN) companion in a close binary system. The hyper-critical accretion induces allows the NS to reach the critical mass \cite{Belvedere2012} finally collapsing to a black hole (BH). In the specific case of GRB 970828, this episode is clearly identified, see Fig. \ref{fig:4b.0}. The observed hard X-ray emission is composed of a thermal spectrum plus a power-law component, both evolving in time. The presence of an evolving thermal component allows the determination of the time decay of the blackbody temperature $kT$ (from 80 to 25
 keV), in the rest-frame time of 20 s, leading to the estimate of the emitter radius between 5000 and 25000 km. 
\item The ``Episode 2'', corresponding to the observations of the GRB, is related to the collapse of the NS into a BH. The characteristic parameters of the GRB 970828 are the Lorentz Gamma factor of $\Gamma \approx 150$, the baryon load $B = 7 \times 10^{-3}$ and a large circumburst density of the order of 10$^3$ particles/cm$^3$.
\item The ``Episode 3'', in soft X-rays, occurs when the prompt emission from the GRB fades away and an additional component, discovered by Swift XRT \cite{Zhang2006,Nousek2006} emerges. It has been shown \cite{Pisani2013} that this component, in energetic ($E_{iso} \geq 10^{52}$ erg) BdHNe, when referred to the rest-frame of the source, follows a standard behavior of the light curve evolution. This emission encompasses the SN shock break out and the expanding SN ejecta ($v/c \approx 0.1$). In the case of GRB 970828, the X-ray emission observed by ASCA and ROSAT perfectly overlap with the common trend observed in BdHN systems and exemplified in Fig. \ref{fig:GS}.
\item The ``Episode 4'' is represented by the observations of the optical emission of the SN, which has been observed in some BdHN sources, with $z \leq 0.9$, \cite[GRB 090618, GRB 060729, GRB 091127, GRB 111228, GRB 080319B, see e.g.][]{Pisani2013}. It is generally hard to detect a SN at $z$=0.9578 and in the case of GRB 970828 is even more difficult due to the very large presence of circumburst material, which has also hampered the observations of the optical afterglow, making this source a 'dark' GRB. 
\end{itemize}

The presence of an evolving thermal component in the first 20 s of the emission of GRB 970828, using BATSE data, has been indicated by \cite{Peer2007}, where they have considered the emission in the first 20 s of Episode 1. They then have fitted the evolution of the temperature $kT$ and the ratio $\mathcal{R} = (Flux_{BB}/\sigma T_{obs}^4)^{1/2}$, where $Flux_{BB}$ is the observed flux of the instantaneous blackbody, $\sigma$ the Stefan constant and $T_{obs}$ the observed temperature, whose evolution in time is fitted with a broken power-law function, see Fig. \ref{fig:4b.0}. In their theoretical interpretation, this thermal emission was associated to the photospheric GRB emission of a relativistic expanding fireball \cite{Meszaros2002}, and they inferred a bulk Lorentz Gamma factor of the expanding plasma, $\Gamma \approx (305 \pm 28) Y_0^{1/4}$, with $Y_0 \geq 1$ the  ratio between the entire fireball energy and the energy emitted in $\gamma$-rays \cite{Meszaros2002}. Since the fireball photospheric radius is given by $r_{ph} = (\mathcal{R} \, D \, \Gamma) / (1.06 (1+z)^2)$, with $D$ the luminosity distance of the GRB, they obtain for $r_{ph}$ the value of 2.7 $\times$ 10$^{11}$ $Y_0^{1/4}$. In \cite{Peer2007}, the authors have also attributed the remaining GRB emission to an unspecified engine activity and neglected all data after 20 s. In their own words, ``we neglect here late-time episodes of engine activity that occur after $\sim$ 25 and $\sim$ 60 s in this burst''. As we will show in the following of this article, we notice the presence of a thermal component in the first 40s, and we attribute it to a non-relativistic initial expansion with radius evolving from $2 \times 10^9$ to $3 \times 10^{10}$ cm,  see Fig. \ref{fig:4b.1}. In addition, we identify the GRB emission between t$_0$+50 s and t$_0$+90 s and the third episode between 10$^4$-10$^6$ s.

In Section 2 we give a summary of the observations of GRB 970828 and describe our data analysis. We proceed in Section 3 to the description of Episode 1, with the details of the expanding black body emitter, the analysis of the non-thermal component and its interpretation in the BdHN paradigm.
In Section 4 we describe Episode 2, the authentic GRB emission. It is well explained in the context of the fireshell scenario, see e.g. \cite{Ruffini2007b} for a complete review of the model. In Section 5 we describe Episode 3, pointing out the clear overlapping of the observed late X-ray data within the theoretical expectation of a BdHN member. In Section 6 we discuss about the theoretically expected SN emission, not observed due to the large circumburst medium. Conclusions are given in the last Section.

\section{Data Analysis}\label{sec:2}

GRB 970828 was discovered with the All-Sky Monitor (ASM) on board the Rossi X-Ray Timing Explorer (RXTE) on 1997 August 28th \cite{IAUC6728}.
Within 3.6 hr the RXTE/PCA scanned the region of the sky around the error box of the ASM burst and detected a weak X-ray source \cite{IAUC6727,IAUC6729}.
GRB 970828 was also observed by the Burst and Transient Source Experiment (BATSE) and the GRB experiment on Ulysses \cite{IAUC6728}. 
The BATSE-LAD light curve is characterized by two main emission phenomena, see Fig. \ref{fig:4b.0}: the first lasts about 40 s and is well described by two main pulses, the second one is more irregular, being composed by several sharp pulses, lasting other 40 s.

The X-ray afterglow was discovered by the ASCA satellite 1.17 days after the GRB trigger \cite{Murakami1997}. The X-ray afterglow observations continued up to 7-10 days from the burst detection.
The optical observations, which started about 4 hr after the burst, did not report any possible optical afterglow for GRB 970828 up to $R$=23.8 \cite{Groot1998}. However, the observations at radio wavelengths of the burst position, 3.5 hrs after the initial burst, succeeded in identifying a source at a good significance level of 4.5 $\sigma$ \cite{Djorgovski2001} inside the ROSAT error circle (10''). The following deep searches for a possible optical counterpart of this radio source led to the identification of an interacting system of faint galaxies, successively recognized as the host galaxy of GRB 970828. The spectroscopic observations of the brightest of this system of galaxies led to the identification of their redshift, being $z$=0.9578. The lack of an optical transient associated with the afterglow of GRB 970828 can be explained as due to the presence of strong absorption, due to dusty clouds in the burst site environment, whose presence does not affect the X-ray and the radio observations of the GRB afterglow. The absence of an optical afterglow \cite{Groot1998}, together with the large intrinsic absorption column detected in the ASCA X-ray data \cite{Yoshida2001} and the contemporary detection in radio-wavelengths of the GRB afterglow, imply a very large value for the circum-burst medium (CBM); the variable absorption might be an indication of a strong inhomogeneous CBM distribution.

To analyze in detail this GRB, we have considered the observations of the BATSE-LAD detector, which observed GRB 970828 in the 25-1900 keV energy range, and then we have reduced the data by using the RMFIT software package.
For the spectral analysis we have considered the High Energy Resolution Burst (HERB) data, which consist of 128 separate high energy resolution spectra stored during the burst emission.
The light curve, shown in Fig. \ref{fig:4b.0}, was obtained by using the Medium Energy Resolution (MER) data, which consist of 4.096 16-channel spectra summed from triggered detectors. 

\section{The Episode 1}

\subsection{The onset of the Supernova and the hyper-critical accretion}

In analogy to the cases of GRB 090618 \cite{Izzo2012}, we analyze here the first emission episode in GRB 970828 to seek for a thermal signature. We have rebinned the light curve assuming a signal-to-noise ratio for each time bin of 20. This large value of counts per bin allows us to consider a gaussian distribution for the photons in each bin, so in the following we will use a $\chi^2$ statistic.

As often done in GRB analysis, we first perform a time-integrated spectral analysis of the first 40 s of emission, which corresponds to the Episode 1, to identify the best-fit model and the possible presence of thermal features. We make use of different spectral models, see Table \ref{tab:4b.1}, to determine the best-fit function. We also check if nested models really improve the best-fit, as in the case of models with an extra power-law component.
We find that the best-fit corresponds to a double blackbody model with an extra power-law component. The check between the Band and the double black body plus power law is minimal ($P_{val} = 9 \%$) but with this last model we note an improvement of the best fit at high energies.

\begin{table*}
\tiny
\centering
\caption{Spectral analysis (25 keV - 1.94 MeV) of the first 40 s of emission in GRB 970828. The following symbols represent: $\ast$ temperature (keV) of the second black body; $\dag$ normalization of the power law component in units of ph cm$^{-2}$ s$^{-1}$ keV$^{-1}$.} 
\vspace{5mm}
\label{tab:4b.1} 
\begin{tabular}{l c c c c c c c c }
\hline\hline
Spectral & $\alpha$ ($\gamma$) & $\beta$ & $\gamma_{ext}$ & $E_{peak}$ & $kT$ &  $\chi^2/DOF $\\
model & & &  & (keV) & (keV)  &  \\ 
\hline 
Power Law & -1.38 $\pm$ 0.01 & - & - & - & -  & 6228.1/115 \\
Cut-off PL & -0.77 $\pm$ 0.02 & - & - & 465.4 $\pm$ 10.6 & - & 203.83/114 \\
Band & -0.60 $\pm$ 0.03 & -2.15 $\pm$ 0.05 & - & 360.5 $\pm$ 12.5  & - & 106.48/113 \\
Band+PL & -0.41 $\pm$ 0.15 & -2.41 $\pm$ 0.33 & -1.47 $\pm$ 0.17 & 335.8 $\pm$ 17.6  & - & 104.12/111 \\
cutoff + PL & -0.47 $\pm$ 0.17 & - & -1.28$\pm$ 0.16& 338.7$\pm$17.9 & - &  104.28/112 \\
BB + po & -  & - & -1.50 $\pm$ 0.01 & - & 63.71 $\pm$ 0.92 &  228.09/113 \\
BB + BB + po & -1.53$\pm$0.17 & - & 0.010 $\pm$ 0.001  $^{\dag}$ & 40.01$\pm$ 2.05 $^{\ast}$ & 106.8 $\pm$ 6.3 & 101.78/111\\
\hline
\end{tabular}
\end{table*}

\begin{figure*}
\centering      
\includegraphics[width=0.49\hsize,clip]{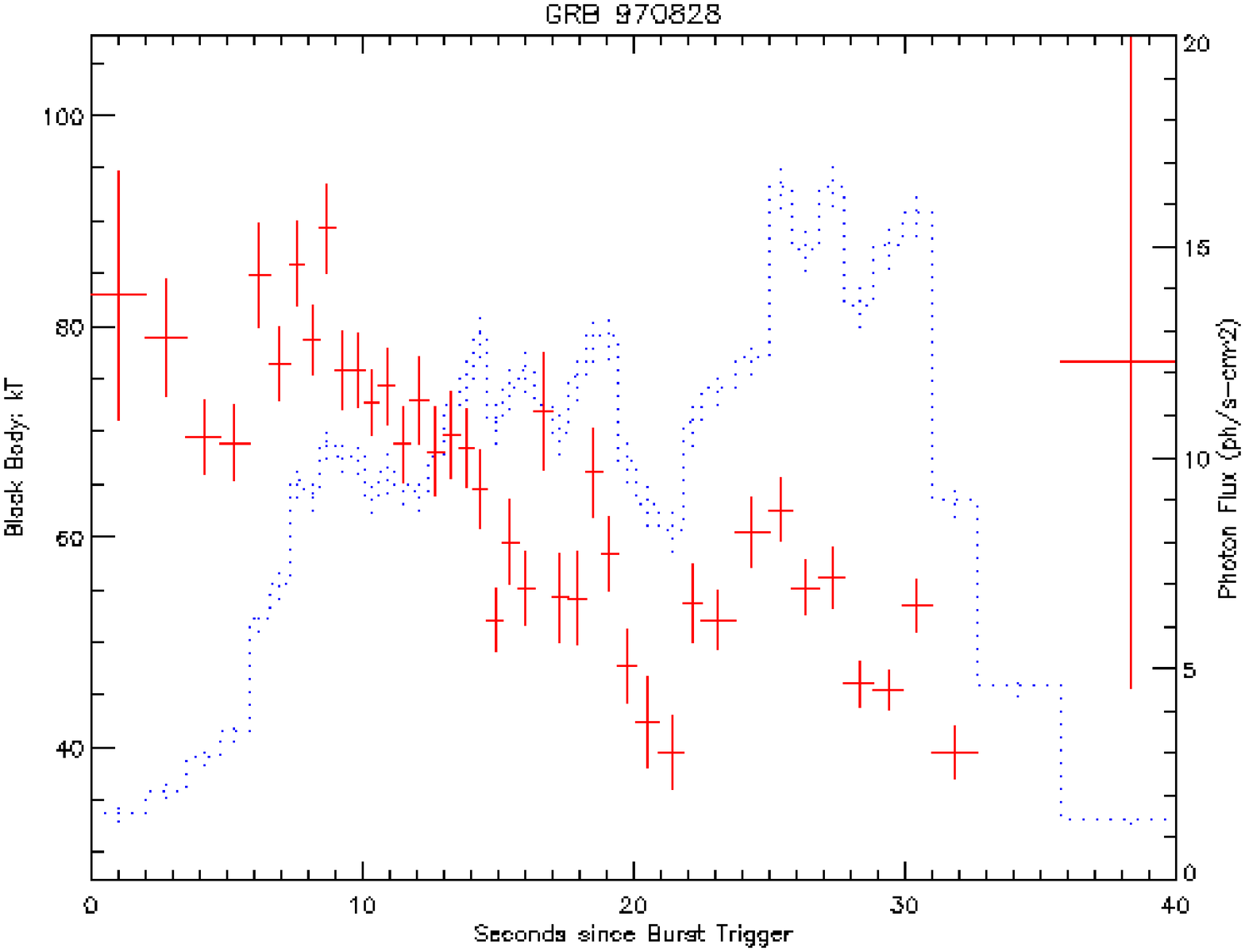}
\includegraphics[width=0.49\hsize,clip]{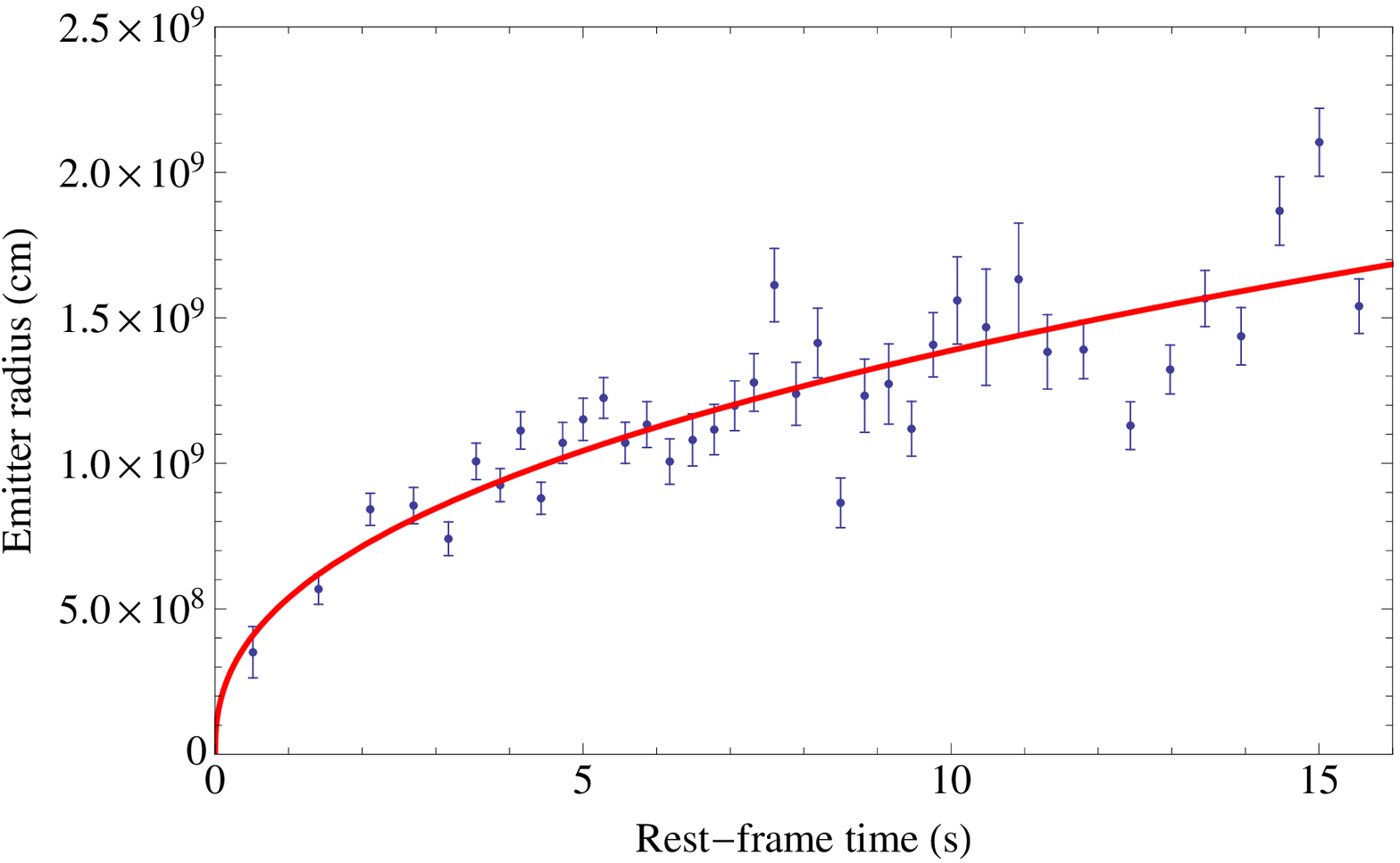}
\caption{(\textit{Left panel}) The evolution of the temperature $kT$ (red crosses) as obtained from a time-resolved spectral analysis of the first 40 s of emission of GRB 970828. The light curve of the first episode (blue dots) is shown in background. (\textit{Right panel}) Evolution of the rest-frame radius of the first episode of GRB 970828. The solid line corresponds to the best fit of this dataset with a power-law function r $\propto$ r$^{\delta}$, with $\delta = 0.41 \pm 0.04$.}
\label{fig:4b.1}
\end{figure*}

It has already been emphasized that the integrated spectral analysis often misses the nature of the physical components and also the nature of the underlying physical mechanisms. We perform therefore a time-resolved spectral analysis to determine the existence and the evolution of a thermal component.
We find that the double blackbody model observed in the time-integrated spectrum can be explained by the presence of an instantaneous single blackbody with a temperature $kT$ varying in intensity and time, showing a double decay trend. We note that the timing of these trends corresponds to the two main spikes in the observed light curve of this first episode, see Fig. \ref{fig:4b.1}.
We have then analyzed this characteristic evolution of the blackbody in both time intervals, corresponding each one to an observed decay trend of the temperature.
From the observed flux of the blackbody component $\phi_{BB,obs}$ for each interval, we obtain the evolution of the emitter radius in the rest-frame:
\begin{equation}\label{eq:no1}
r_{em} = \left(\frac{\phi_{BB,obs}}{\sigma T_{obs}^4}\right)^{1/2} \frac{D}{(1+z)^2}
\end{equation}
whose  evolution  is shown in Fig. \ref{fig:4b.1}. 
It is very interesting that the radius monotonically increases, without showing an analog double trend which is observed for the temperature, see Fig. \ref{fig:4b.1}.
The global evolution of the emitter radius is well-described with a power-law function $r = \alpha t^{\delta}$ and a best fit of the data provides for the $\delta$ = 0.41 $\pm$ 0.04 and $\alpha$ = (5.38 $\pm$ 0.52) 10$^8$ cm, with an $R^2$ statistic value of 0.98, see Fig. \ref{fig:4b.1}.

It is appropriate to discuss the power law component observed in the time resolved spectra. In BdHNe, the tight geometry of the binary system implies that as the external layers of the SN core starts to expand, an hyper-critical accretion phenomenon is induced onto the NS companion.

\subsection{Binary progenitor and binary-driven hypercritical accretion}

The first estimates of the IGC process \cite{Rueda2012,Izzo2012}
were based on a simplified model of the binary parameters and the
Bondi-Hoyle-Lyttleton accretion formalism \cite{1939PCPS...35..405H,1944MNRAS.104..273B,1952MNRAS.112..195B}. The following discussion is based on the more recent and accurate results presented in \cite{FRR2014}, in which the collapsing CO cores leading to SN Ic are simulated to calculate realistic profiles for the density and ejection velocity of the SN outer layers. The hydrodynamic evolution of the accreting material falling into the Bondi-Hoyle accretion region is also computed from numerical simulations all the way up to its
incorporation onto the NS surface.

The hypercritical accretion onto the NS from the SN ejecta can be estimated from the Bondi-Hoyle-Lyttleton
formula 
\begin{equation}\label{eq:Mdot}
\dot{M}_{\rm BHL}=4 \pi r_{\rm BHL}^2 \rho (v^2 +c_s^2)^{1/2}\, ,
\end{equation}
where $\rho$ is the SN ejecta density, $v$ is the ejecta velocity in the rest-frame of the NS, which includes a
component from the ejecta velocity, $v_{\rm ej}$, and another component from the orbital velocity of the NS, $v_{\rm orb}$; $c_s$
is the SN ejecta sound speed, and $r_{\rm BHL}$ is the Bondi radius
\begin{equation}\label{eq:rbhl}
r_{\rm BHL}=\frac{G M_{\rm NS}}{v^2 +c_s^2}\, ,
\end{equation}
being $G$ the gravitational constant and $M_{\rm NS}$ is the NS mass. The conditions of the binary system are such that both the velocity components, $v_{\rm orb}, v_{\rm ej}$, are typically much higher than the sound speed. The ejecta velocity as a function of time is determined by the explosion energy
and the nature of the SN explosion. The orbital velocity depends upon
the orbital separation, which in turn depends upon the radius of the
CO core and the binary interactions prior to the explosion of the CO core. The effect of the NS magnetic field is negligible in this process \citep{1996ApJ...460..801F,Rueda2012}: for a neutron star with surface magnetic field $B=10^{12}$~G, mass $M_{\rm NS}=1.4~M_\odot$, and radius $r_{\rm NS}=10^6$~cm, one has that for accretion rates $\dot{M}>2.6\times 10^{-8}~M_\odot$~s$^{-1}=0.8~M_\odot$~yr$^{-1}$, the Alfv\'en magnetospheric radius satisfies $R_A=[B^2 R^6/(\dot{M} \sqrt{2 G M_{\rm NS}})]^{2/7}<r_{\rm NS}$.

The evolution of the SN ejecta density near the NS companion depends on the SN explosion and the structure of the progenitor
just prior to collapse. The compactness of the CO core is such that there is no Roche lobe overflow prior to the SN explosion. The Roche lobe radius can be computed from \cite{1983ApJ...268..368E}, $R_{\rm L,CO}/a\approx 0.49 q^{2/3}/[0.6 q^{2/3} + \ln(1 + q^{1/3})]$, where $q=M_{\rm CO}/M_{\rm NS}$. For a CO core progenitor $M_{\rm CO}\approx 5~M_\odot$, $R_{\rm CO}\approx 3\times 10^9$~cm, no Roche lobe overflow occurs for binary periods $P\geq 2$~min, or binary separations $a\geq 6\times 10^9$~cm, assuming a NS companion mass $M_{\rm NS}\geq 1.4~M_\odot$.

In order to derive the accretion onto the NS, the explosion has to be modeled. We have recently performed the numerical simulations following two different approaches \cite{FRR2014}: the first assuming a homologous outflow with a set explosion energy and a second approach following the collapse, bounce, and explosion of a 20$M_\odot$ (zero-age main sequence mass) progenitor.  The calculation uses a 1D core-collapse code \citep{1999ApJ...516..892F} to follow the collapse and bounce and then injects energy just above the proto-NS to drive different SN explosions mimicking the
convective-engine paradigm. With this progenitor and explosion, we produce the density and velocity evolution history at the position of the Bondi-Hoyle surface of the NS companion. 

Under the above conditions, we have found from our numerical simulations in \cite{FRR2014} that hypercritical accretion rates of up to $10^{-2}\,M_\odot/$s occur in these systems. This infall rate is well above the critical Eddington rate. The Eddington accretion limit, or critical accretion rate makes a series of assumptions: 1) the potential energy gained by the accreting material is released in the form of photons which exert pressure finally reducing the accretion rate, 2) the inflowing material and outflowing radiation is spherically symmetric, 3) the photons are not trapped in the flow and can deposit momentum to the inflowing material, and 4) the opacity is dominated by electron scattering. However, many of these assumptions break down in the IGC scenario, allowing hypercritical accretion rates.

It can be shown that the photons for the hypercritical accretion rates in the IGC are trapped in the flow. \citet{1989ApJ...346..847C} derived the trapping radius where photons emitted diffuse outward at a slower velocity than infalling material flows inward:
\begin{equation}
r_{\rm trapping} = min [(\dot{M}_{\rm BHL} \kappa)/(4 \pi c),r_{\rm BHL}]
\end{equation}
where $\kappa$ is the opacity (in cm$^2$~g$^{-1}$) and $c$ is the speed of light. If the trapping radius is near or equal to the Bondi radius, the photons are trapped in the flow and the Eddington limit does not apply.  We estimate for our CO core a Rosseland mean opacity roughly $5\times 10^3$~cm$^2$~g$^{-1}$, a factor $\sim 10^4$ higher than electron scattering. Combined with our high accretion rates, it is clear that the Eddington limit does not apply in this scenario and hypercritical accretion must occur. 

The inflowing material shocks as it piles up onto the NS producing an atmosphere on top of the NS which, by compression, becomes sufficiently hot to emit neutrinos \cite{1972SvA....16..209Z,1989ApJ...346..847C,1991ApJ...376..234H,1996ApJ...460..801F}. The neutrinos have become then crucial in cooling the infalling material, allowing its incorporation into the NS \cite{1973PhRvL..31.1362R,1996ApJ...460..801F,2009ApJ...699..409F}. We compute the neutrino emission following \cite{1996ApJ...460..801F,2009ApJ...699..409F}. We thus take into account $e^-$ and $e^+$ capture by free protons and neutrons, and pair and plasma $\nu \bar{\nu}$ creation; $\nu$ absorption processes include $\nu_e$ capture by free neutrons, $\bar{\nu}_e$ by free protons, and $\nu\bar{\nu}$ annihilation. $\nu$ scattering includes $e^-$ and $e^+$ scattering off $\nu$ and neutral current opacities by nuclei. The three species $\nu_{e,\mu,\tau}$ are tracked separately by the transport algorithm.

As material piles up, the accretion shock moves outward. The accretion shock weakens as it moves out and the entropy jump becomes smaller, producing an unstable atmosphere with respect to Rayleigh-Taylor convection. Previous simulations \cite{2006ApJ...646L.131F,2009ApJ...699..409F} of such instabilities accretion process have shown that they can accelerate above the escape velocity driving outflows from the accreting NS with final velocities approaching the speed of light, causing the ejection of up to 25\% of the accreting material. The entropy of the material at the base of our atmosphere, $S_{\rm bubble}$, is given by \citep{1996ApJ...460..801F}:
\begin{equation}
S_{\rm bubble}=38.7~\left(\frac{M_{\rm NS}}{2M_\odot} \right)^{7/8} \left(\frac{\dot{M}_{\rm
  BHL}}{0.1~M_\odot {\rm s}^{-1}}\right)^{-1/4}\left(\frac{r_{\rm NS}}{10^6~{\rm cm}}\right)^{-3/8}
\end{equation}
$k_B$ per nucleon, where $r_{\rm NS}$ is the radius of the NS.  The corresponding temperature of the bubble, $T_{\rm bubble}$, is:
\begin{equation}
T_{\rm bubble} = 195~S_{\rm bubble}^{-1} \left(\frac{r_{\rm NS}}{10^6~{\rm cm}}\right)^{-1}.
\end{equation}

Under the hypercritical accretion of the IGC, the temperature of the bubble when it begins to rise is $T_{\rm bubble}\sim$5~MeV. If it rises adiabatically, expanding in all dimensions, it drops to 5~keV at a radius of $10^9$~cm, far too cool to observe. However, if it is ejected in a jet, as simulated in \citet{2009ApJ...699..409F}, it expands laterally but not radially, so we have roughly $\rho \propto r^{2}$ and $T \propto r^{-2/3}$. In that simplified bubble evolution, the outflow would have a temperature $T_{\rm bubble}\sim 50$~keV at $10^9$~cm and $T_{\rm bubble}\sim 15$~keV at $6\times10^9$~cm. This could explain the temperature and size evolution of the blackbody observed in the Episode 1 of BdHNe. For example, the blackbody observed in Episode 1 of GRB 090618 \citep{Izzo2012} evolves as $T\propto r^{-m}$ with $m=0.75 \pm 0.09$, in agreement with this simplified theoretical estimate. For the present case of GRB 970828, the fully lateral bubble evolution do not match perfectly, implying that the above simplified picture needs further refinement and/or the presence of other mechanisms. We are currently deepening our analysis of the possible explanation of the thermal emission observed in Episode 1 of BdHNe as based on convective instabilities in the hypercritical accretion process, and the results will be presented elsewhere.

Concerning the power-law component observed in the luminosity of Episode 1 in addition to the blackbody one, we advance the possibility that such a high-energy emission could come from the angular momentum of the binary system as follows.

The angular momentum per unit mass accreting by the NS can be estimated as
\begin{equation}
j_{\rm acc}\approx \frac{1}{2} \omega_{\rm orb} r_{B}^2,
\end{equation}
where $\omega_{\rm orb}=v_{\rm orb}/a$ is the orbital angular velocity, $v_{\rm orb}=(G M_T/a)^{1/2}$ is the orbital velocity, $a$ the separation distance of the binary components, $M_T = M_{\rm CO}+M_{\rm NS}$ is the total mass of the binary. $r_B$ is the Bondi capture radius. From our numerical simulations, we know that when the neutron star reaches the critical mass, the inequality $v_{\rm ej}\ll v_{\rm orb}$ is satisfied, so we can approximate Eq.~(\ref{eq:rbhl}) as
\begin{equation}
r_{\rm BHL}\approx \frac{2 G M_{\rm NS}}{v^2_{\rm orb}}\to \frac{2 G M_{\rm crit}}{v^2_{\rm orb}} = \frac{2 G M_{\rm BH}}{v^2_{\rm orb}},
\end{equation}
where $M_{\rm BH}=M_{\rm crit}$, is the mass of the newly-formed black hole, so it equals $M_{\rm crit}$, the critical mass of the NS.

The black hole can gain angular momentum up to it reaches the maximal value allowed by the Kerr solution
\begin{equation}
j_{\rm maxBH}=\frac{G M_{\rm BH}}{c}.
\end{equation}

\begin{figure}[!hbtp]
\centering
\includegraphics[width=0.49\hsize,clip]{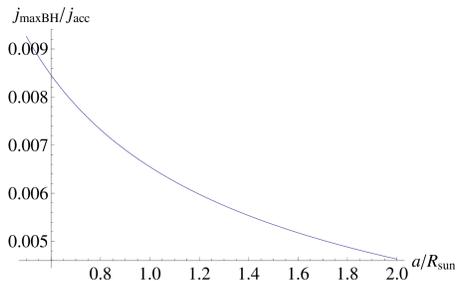}
\caption{Maximal black hole to accretion angular momentum ratio, $j_{\rm maxBH}/j_{\rm acc}$, as a function of the binary separation in units of solar radius. Here for simplicity we have used $M_{\rm CO}\approx 6\,M_\odot$ and $M_{\rm BH}\approx 3\,M_\odot$.}\label{fig:jratio}
\end{figure}

Therefore we have (see Fig.~\ref{fig:jratio})
\begin{equation}\label{eq:jratio}
\frac{j_{\rm maxBH}}{j_{\rm acc}}=\frac{1}{2}\frac{M_T}{M_{\rm BH}} \sqrt{\frac{G M_T}{c^2 a}}=\frac{1}{2}\left(1+\frac{M_{\rm CO}}{M_{\rm BH}}\right) \sqrt{\frac{G (M_{\rm CO}+M_{\rm BH})}{c^2 a}}.
\end{equation}

It becomes then clear from the above first simplified estimate that the angular momentum carried out by the accreted material highly exceed the maximal angular momentum that the newly-born black hole can support, and therefore angular momentum dissipation, very likely in form of collimated emission, is likely to occur. We are currently performing numerical simulations of this process in order to assess the validity and accuracy of these first order of magnitude estimates.

\subsection{A possible explanation for the non-thermal component and the compactness problem}

It is well known \cite[see][]{Piran99} that most of GRBs emit a large fraction of observed high-energy photons ($E \gg1$ MeV) which can interact with low-energy photons to produce electron-positron pairs via $\gamma\gamma\rightarrow e^+e^-$ in a compact region with radius R that, with a naive estimate, can be considered $<c\delta t$. This would imply an optical depth  $\tau >> 1$, but we know that GRB spectra are non-thermal, so we are in presence of a paradox. This issue can be solved assuming a relativistic expansion of the emitting source, with Lorentz factor $\Gamma \gg 1$ \cite{Ruderman,Piran99}.. In this case, in fact, we would have $R < 2\gamma^2 c dt$ and consequently a decrease of the estimated optical depth \cite{WoodsLoeb,LitSari2001,ZhangPeer}.

The observed high-energy photon spectrum is often modeled by a single power-law $K E^{-\gamma}$, with $E_{min} < E < E_{max}$ and power-law index $\gamma$.
The energy $E_{max}$ is the highest observed photon energy.
In the frame of the emitting material, where the photons are assumed to be isotropic, a photon with energy $E^\prime$ can annihilate a second photon with energy $E_{th}^\prime$, yielding an electron-positron pair.
The threshold for this process is described by
\begin{equation}
\label{thresh}
E^\prime E_{th}^\prime \geq(m_ec^2)^2\ ,
\end{equation}
where $m_e$ is the electron mass. 
If the source is moving toward the observer with a Lorentz factor $\Gamma$, then the photons previously analyzed have detected energy of $E=\Gamma E^\prime/(1+z)$ and $E_{th}\geq\Gamma E^\prime_{th}/(1+z)$, respectively.
Therefore in the observer frame photons with energy $E_{max}$ annihilate only with other photons having energy $E_{max,th}=[\Gamma m_ec^2/(1+z)]^2 E_{max}^{-1}$.

Although the observed high-energy photon spectrum is a power-law up to 4 MeV in the rest frame of the burst, there is no observational evidence for the presence of a cut-off due to the $e^+e^-$-pair creation. Therefore we can estimate the minimum Lorentz factor of the non-thermal component allowed by the observations from the maximum energy observed in the first episode $E_{max} \sim 2 MeV$. From this assumption, it is straightforward to impose that the threshold energy $E_{max,th}$ for the pair creation process has to be $E_{max,th} < E_{max}$ \cite[see e.g. case (III) in][]{Gupta}.
It follows then a lower limit on the Lorentz factor from the observed energy $E_{max}$
\begin{equation}
\Gamma_{min} \geq \frac{E_{max}}{m_e c^2}\left(1+z\right)\ .
\label{uplim}
\end{equation}
We can identify $E_{max}$ with the cut-off energy of the spectrum $E_c$, but for the moment we treat them as different energies.
Following the considerations in \cite{Gupta}, we have calculated the averaged number of photons interacting with $E_{max}$ from $E_{max,th}$ to $E_c\geq E_{max}$ on the cross-section of the process integrated over all the angles $\theta$
\begin{equation}
\langle \sigma N_{max,th}\rangle = 4\pi d_z^2 \Delta t \int_{E_{max,th}}^{+ \infty}{K E^{-\gamma}dE} \int_{1}^{\frac{E_{max}E}{(m_ec^2)^2}}{\frac{3}{16}\sigma_T s ds} = \frac{2E_{max,th}^{1-\gamma}}{\xi}
\end{equation}
and we have correspondingly evaluated the optical depth
\begin{equation}
\tau_{\gamma\gamma} = \frac{\langle \sigma N_{max,th}\rangle}{4\pi (\Gamma^2 c \Delta t)^2}\ ,
\label{tautautaumod}
\end{equation}
by defining the following quantities
$$d_z=\frac{D}{1+z}\hspace{0.2cm},\hspace{0.2cm} s=\frac{E_{max}E(1-\cos\theta)}{2(m_ec^2)^2} \hspace{0.2cm},\hspace{0.2cm}\xi\equiv\left[\frac{3\pi\sigma_T d_z^2 K\Delta t}{4(\gamma^2-1)}\right]^{-1}\ ,$$ 
and using the Thomson cross-section $\sigma_T$.
The condition $\tau_{\gamma\gamma} < 1$ yields to a lower limit on the Lorentz $\Gamma$ factor.
We have applied these considerations to non-thermal spectrum of the first episode of GRB 970828, and considered for $\Delta$ t in Eq. \ref{tautautaumod} the whole duration of the first episode in GRB 970828.
Therefore we have calculated an averaged lower limit on the Lorentz factor, i.e. $\Gamma_{min}=77$ for the whole first episode.%

Therefore, a relativistic outflow of the accretion process of the SN onto the companion NS, can explain the origin of the power-law high energy component observed in Episode 1.

\section{The Episode 2 : the GRB emission}

Turning now to the second emission episode, we have computed the isotropic energies emitted in this episode, by considering a Band model as the best fit for the observed integrated spectra:  $E_{iso,2nd} = 1.6 \times 10^{53}$ erg.
In what follows we explain this second emission episode of GRB 970828 as a single canonical GRB emission in the context of the Fireshell scenario.

In this model \cite{Ruffini2001L107,Ruffini2001c}, a GRB originates from an optically thick $e^+e^-$-plasma created in the process of vacuum polarization, during the process of gravitational collapse leading to a Kerr-Newman black hole \cite{Christodoulou,Damour}. The dynamics of this expanding plasma is described by its total energy $E_{tot}^{e^+e^-}$, the baryon load $B = M_B c^2/E_{tot}^{e^+e^-}$ and the circumburst medium (CBM) distribution around the burst site. The GRB light curve emission is characterized by a first brief emission, named the proper GRB or P-GRB, originating in the process of the transparency emission of the $e^+e^-$-plasma, followed by a multi-wavelength emission due to the collisions of the residual accelerated baryons and leptons with the CBM. This latter emission is assumed in a fully radiative regime. Such a condition is introduced for mathematical simplicity and in order to obtain a lower limit on the CBM density. This condition establishes a necessary link between the CBM inhomogeneities and filamentary distribution \cite{Ruffini2005b} with the observed structures in the $\gamma$ and X-ray light curves in the prompt and early afterglow phase. In the spherically symmetric approximation
 the interaction of the accelerated plasma with the CBM can be described by the matter density distribution $n_{CBM}$ around the burst site and the fireshell surface filling factor $\mathcal{R} = A_{eff}/A_{vis}$, which is the ratio between the effective emitting area and the total one \cite{Ruffini2005}. The spectral energy distribution in the comoving frame of the shell is well-described by a ``modified'' thermal emission model \cite{Patricelli2011}, which differs from a classical blackbody model by the presence of a tail in the low-energy range.

\begin{figure}
\centering
\includegraphics[width=0.42\hsize,clip]{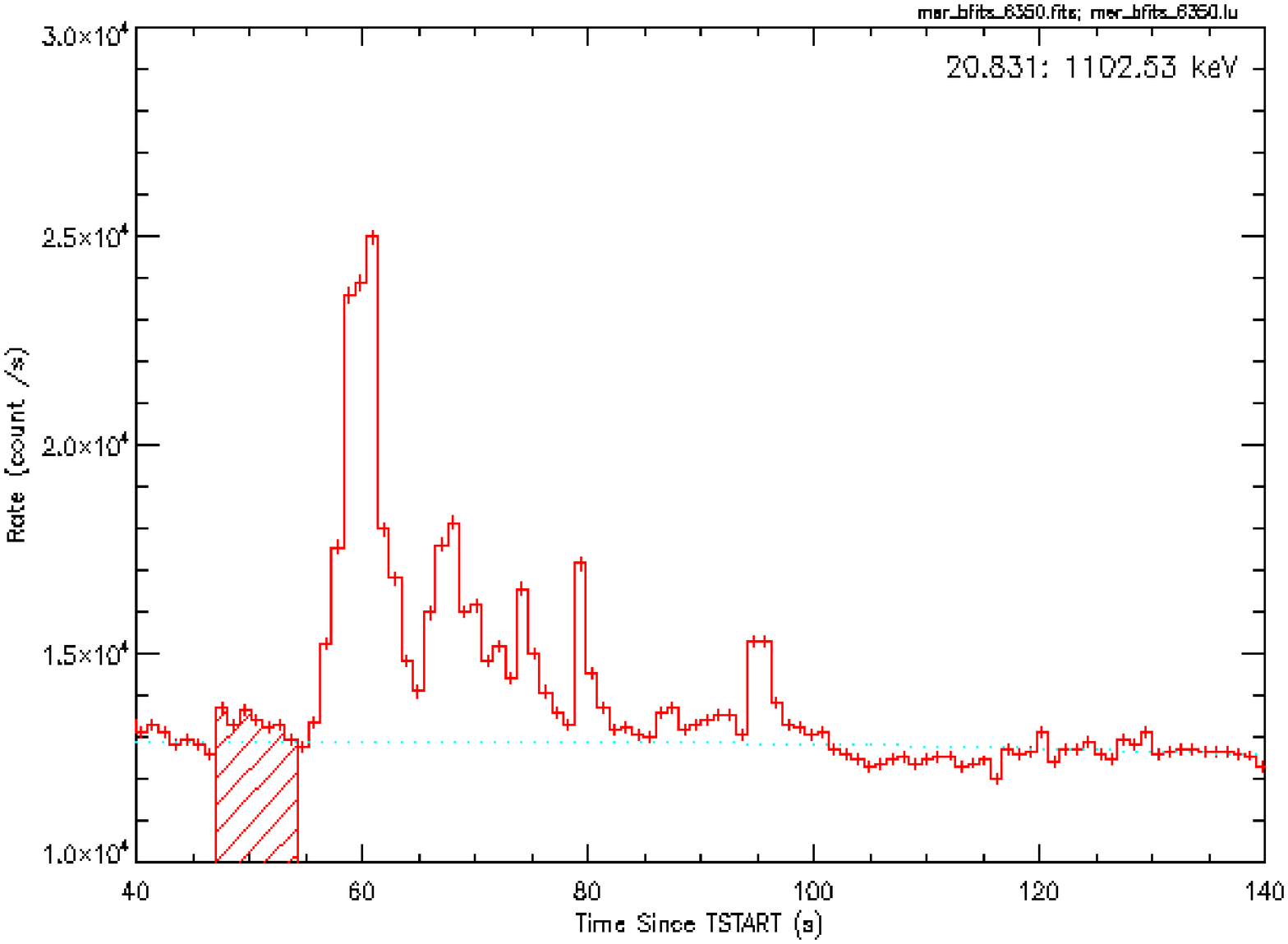}
\includegraphics[width=0.45\hsize,clip]{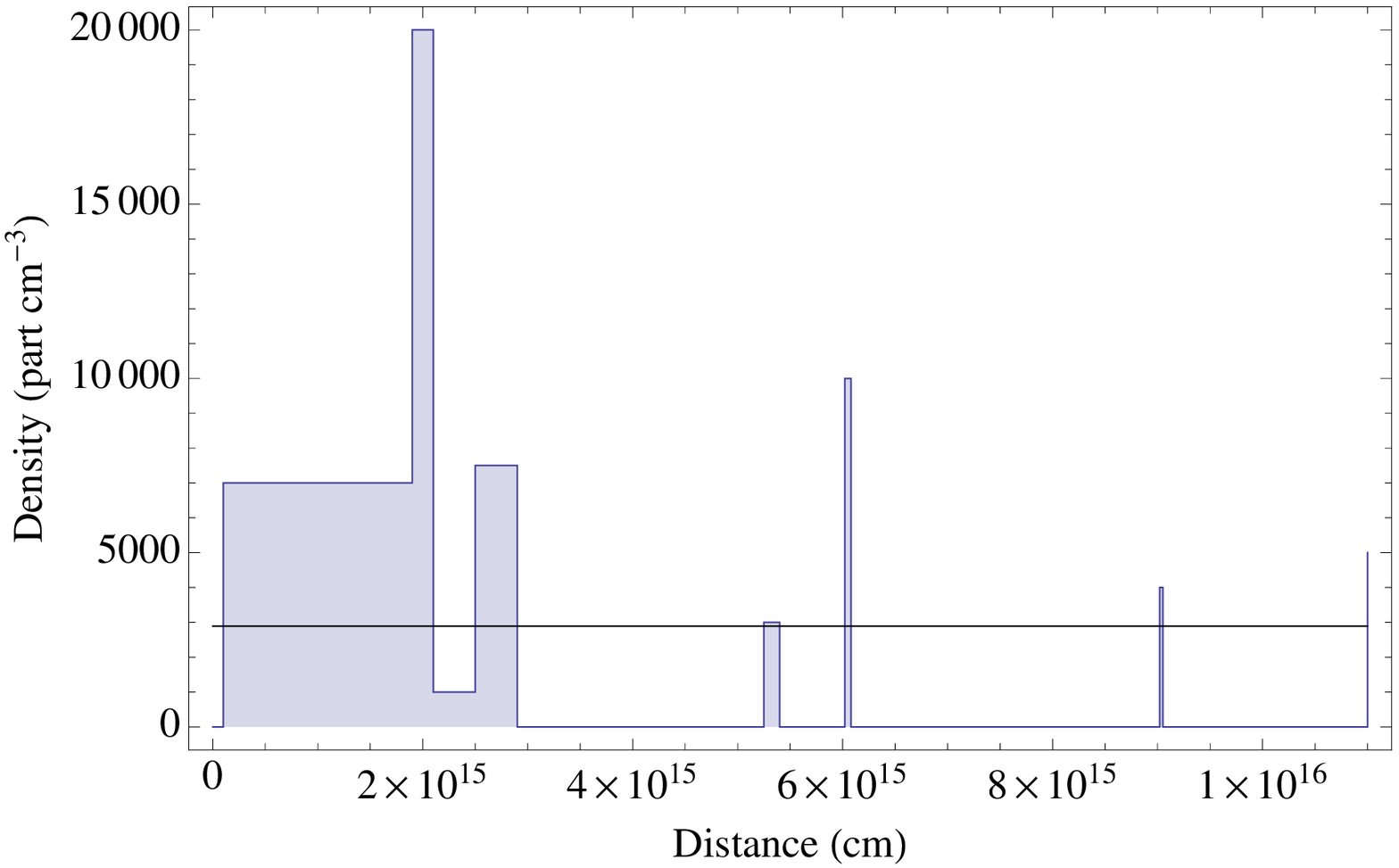}\\
\caption{ \textit{Left panel:} Light curve of the second episode in GRB 970828. The dashed region represents the P-GRB emission.  \textit{Right panel:} The radial CBM density distribution for GRB 970828. The characteristic masses of each cloud are on the order of $\sim$ 10$^{22}$ g and 10$^{15}$ cm in radii. The black line corresponds to the average value for the particle density. }
\label{fig:4b.3b}
\end{figure}

In this context, to simulate the second episode of GRB 970828, which is the actual GRB emission, we need to identify the P-GRB signature in the early second episode light curve.
From the identification of the P-GRB thermal signature, and the consequent determination of the energy emitted at transparency, we can obtain the value of the baryon load $B$ assuming that the total energy of the $e^+e^-$-plasma is given by the isotropic energy $E_{iso}$ observed for the second episode of GRB 970828, as it was done for the second episode in GRB 090618, see e.g. \cite{Izzo2012}.
We have then started  to seek for a possible thermal signature attributable to the P-GRB emission in the early emission of the second episode.
As it is shown in Fig. \ref{fig:4b.3b}, the early emission of the second episode is characterized by an intense spike, anticipated by a weak emission of 9 s.
Our search for the P-GRB emission is concentrated in this time interval, since from the fireshell theory the expected energy of the P-GRB emission, in case of long GRBs for which the baryon load is in between $10^{-3} - 10^{-2}$, is of the order of 10$^{-2}$ of the prompt emission. The observed fluence (10-1000 keV) in the P-GRB emission, computed from the fit with the power-law function is $S_{obs} = (1.54 \pm 0.10) \times 10^{-6}$ erg/cm$^2$, which corresponds to an isotropic energy of the P-GRB of $E_{iso,PGRB} = 1.46 \times 10^{51}$ erg, which is quantitatively in agreement with the energetic of the P-GRB for this GRB (it is $\approx$ 0.01 $\%$ the total energy of the second episode, the GRB). However, due to the paucity of photons in this time interval, we are not able to put tight constraints, e.g. about a possible observed temperature of the P-GRB.

\begin{table*}
\tiny
\centering
\caption{Spectral analysis (25 keV - 1.94 MeV) of the P-GRB emission in the second episode of GRB 970828.} 
\vspace{5mm}
\label{tab:4b.2} 
\begin{tabular}{l c c c c c}
\hline\hline
Spectral model & $\gamma$ & $\beta$ & $E_{cutoff}(keV)$ & $kT(keV)$ & $\chi^2/DOF $\\ 
\hline 
power-law & -1.18 $\pm$ 0.04 & &  &   & 91.495/115 \\
cutoff PL & -1.15 $\pm$ 0.08 & & 2251 $\pm$ 1800  &  & 91.157/114 \\
BB + PL & -1.16 $\pm$ 0.06 & &  & 69.6 $\pm$ 40.0 & 90.228/113\\
Band & -0.96 $\pm$ 0.44 & -1.23 $\pm$ 0.08 & 958.8 $\pm$ 800.0 & & 90.439/113\\
\hline
\end{tabular}
\end{table*}

\begin{table}
\centering
\tiny{
\caption{Final results of the simulation of GRB 970828 in the fireshell scenario} 
\vspace{5mm}
\label{tab:4b.3} 
\begin{tabular}{l c}
\hline\hline
Parameter & Value \\ 
\hline 
$E_{tot}^{e^+e^-}$ &  (1.60 $\pm$ 0.03) $\times$ 10$^{53}$ erg\\
$B$ &  (7.00 $\pm$ 0.55) $\times$ 10$^{-3}$\\
$\Gamma_0$ & 142.5 $\pm$ 57\\
$kT_{th}$ & (7.4 $\pm$ 1.3) keV\\
$E_{P-GRB,th}$ & (1.46 $\pm$ 0.43) $\times$ 10$^{51}$ erg\\ 
$\langle  n  \rangle $ & $3.4 \times 10^3 \, part/cm^3$\\   
$\delta n/n$ & $10 \, part/cm^3$\\
\hline
\end{tabular}}
\end{table}

With these results, we can estimate the value of the baryon load from the numerical solutions of the fireshell equations of motion.
These solutions for four different values of the total e$^+$e$^-$-plasma energy are shown in the Fig. 4 of \cite{Izzo2012}.
We find that the baryon load is $B = 7 \times 10^{-3}$, which corresponds to a Lorentz gamma factor at transparency $\Gamma = 142.5$.
The GRB emission was simulated with very good approximation by using a density mask characterized by an irregular behavior: all the spikes correspond to spherical clouds with a large particle density $\langle n \rangle  \sim 10^3$ part/cm$^3$, and with radius of the order of $(4 - 8) \times 10^{14}$ cm, see Fig. \ref{fig:4b.3b}.
Considering all the clouds found in our analysis, the average density of the CBM medium is $\langle n \rangle  = 3.4 \times 10^3$ particles/cm$^3$.
The corresponding masses of the blobs are of the order of $10^{24}$ g, in agreement with the clumps found in GRB 090618.

\section{The Episode 3 : the late X-ray afterglow}

The most remarkable confirmation of the BdHN paradigm applied to GRB 970828, comes from the late X-ray afterglow emission. As shown in \cite{Pisani2013}, from the knowledge of the redshift of the source, we can compute the X-ray luminosity light curve in the common rest frame energy range $0.3$--$10$ keV after $\approx$ 10$^4$ s from the initial GRB emission.
However, while in \cite{Pisani2013} the analysis is based on the available X-ray data ($0.3$--$10$ keV) from the \textit{Swift}-XRT detector, GRB 970828 occurred in the pre-\textit{Swift} era. 
Its observational X-ray data are available in the energy range $2$--$10$ keV, since the data were collected by three different satellites: \textit{RXTE}, \textit{ASCA} and \textit{ROSAT}.
To further confirm the progenitor mechanism for GRB 970828, we verify the overlapping of the late X-ray data with the ones of the 'Golden Sample' (GS) sources presented in \cite{Pisani2013}. To this aim, we have computed its luminosity light curve $L_{rf}$ in a common rest-frame energy range $0.3$--$10$ keV.
Since the observed energy band is different ($2$--$10$ keV), the expression for the flux light curve $f_{rf}$ in the $0.3$--$10$ keV rest-frame energy range is not as expressed in Eq.~2 of \cite{Pisani2013}, but it becomes  
\begin{equation}
\label{frfmod}
f_{rf}=f_{obs}\frac{\left(\frac{10}{1+z}\right)^{2-\gamma}-\left(\frac{0.3}{1+z}\right)^{2-\gamma}}{10^{2-\gamma}-2^{2-\gamma}}\ ,
\end{equation}
where $\gamma$ is the photon index of the power-law spectral energy distribution of the X-ray data.
All the other data transformations, reported in \cite{Pisani2013}, remain unchanged.

\begin{figure}
\centering
\includegraphics[width=0.43\hsize,clip]{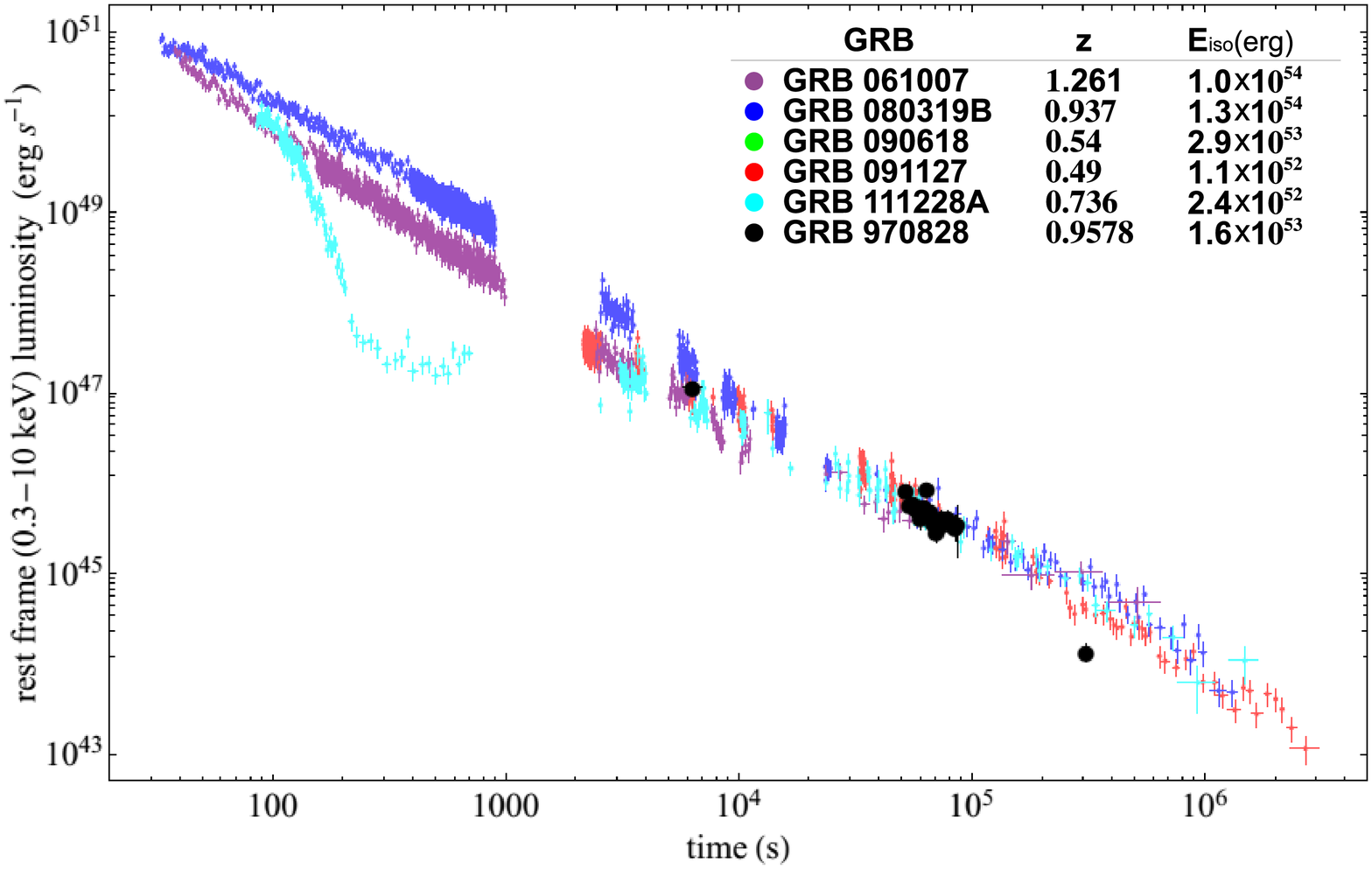}
\includegraphics[width=0.44\hsize,clip]{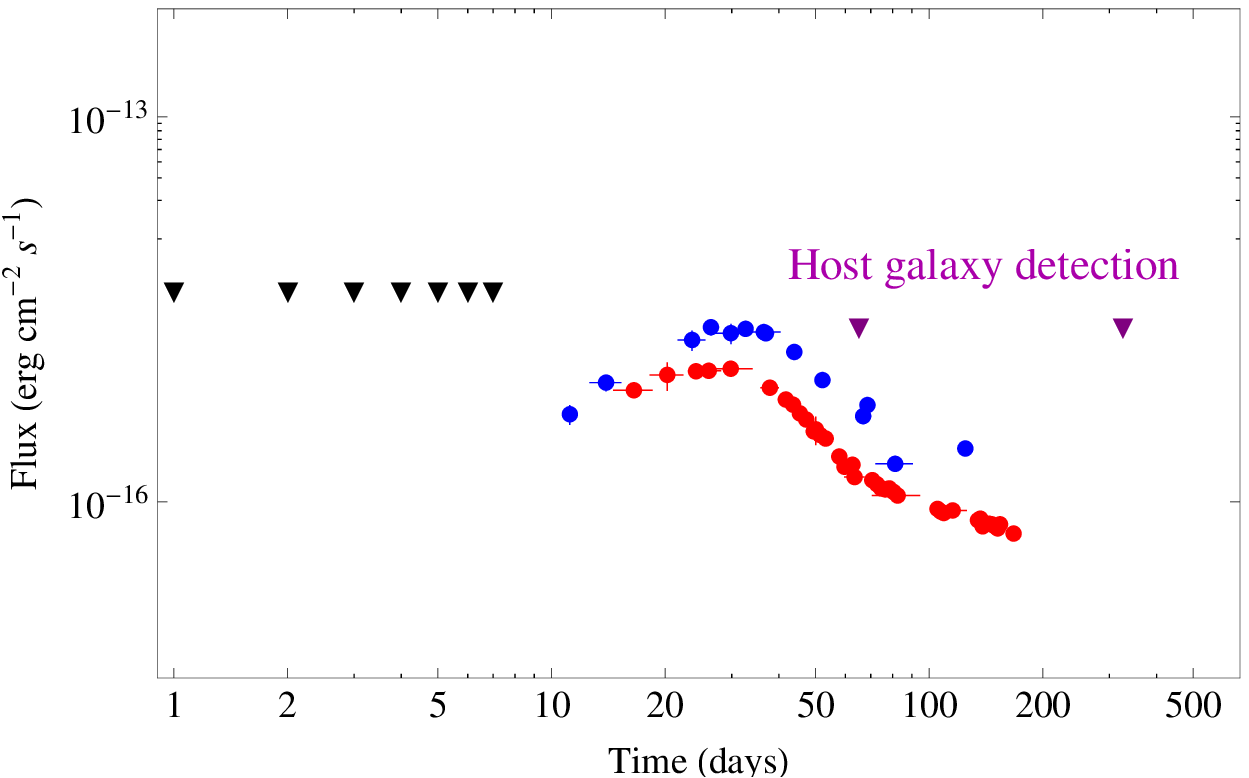}\\
\caption{\textit{Left panel:} The late X-ray (0.3 - 10 keV) light curves of some GRBs presented in \cite{Pisani2013} and of GRB 970828 (black open circles). The overlap of the light curve of GRB 970828 with the members of the BdHN class is clearly evident, confirming that an Induced Gravitational Collapse (IGC) mechanism is operating also in GRB 970828 . \textit{Right panel:} The light curve of the SN associated with GRB 090618 (green data), the $U-$ (blue line) and $R-$band (red line) light curve of SN 1998bw transposed at the redshift of GRB 970828, $z$ = 0.9578, and not corrected for the intrinsic host-galaxy absorption. The purple and cyan line represents the limit given in the deep images by \cite{Djorgovski2001} and \cite{Groot1998} respectively.}
\label{fig:GS}
\end{figure}

We made use in particular of the \textit{RXTE}-PCA observations and \textit{ASCA} data presented in \cite{Yoshida2001}; the averaged photon indexes are taken from the text, for \textit{RXTE}-PCA ($\gamma\sim 2$), and from Tab.~1, for the \textit{ASCA} data, of the same paper.
The last data-point by \textit{ROSAT} is taken from Fig.~7 in \cite{Djorgovski2001}, with a corresponding photon index $\sim 2$; the error on the observed flux is the $25\%$ as indicated for the count rate \cite{IAUC6757}.
We show in Fig. \ref{fig:GS} the late X-ray (0.3 - 10 keV) light curve of GRB 970828 and we compare it with some GRBs of the ``Golden Sample'' presented in \cite{Pisani2013}: GRB 061007, GRB 080319B, GRB 090618, GRB 091127 and GRB 111228A.  The perfect overlap with the late X-ray light curves of BdHN sources confirms the presence of a BdHN mechanism operating also in GRB 970828.

\section{Limits on the Episode 4 : SN-related observations}

The analysis of GRB 090618 \cite{Izzo2012} and GRB 101023 \cite{Penacchioni2012} represents an authentic ``Rosetta Stone'' for the understanding of the GRB-SN phenomenon. The presence of a supernova emission, observed ten days after the burst in the cosmological rest-frame of GRB 090618, was found to have the same luminosity of SN 1998bw \cite{Cano2011}, the SN related to GRB 980425 and which is the prototype of SNe connected with GRBs \cite{DellaValle2011}. We have transposed the data of the ``bump'' Rc-band light curve observed in the optical afterglow of GRB 090618, associated to the presence of an underlying supernova \cite{Cano2011}, to the redshift of GRB 970828. This simple operation concerns only the transformation of the observed flux, under the assumption that the SN has the same intrinsic luminosity. Moreover, we have also transposed the $U$ and $R-$band light curves of SN 1998bw \cite{Galama1998}, which is the prototype of a supernova associated to a GRB. From the $K-$ correction transformation formula, the $U-$band light curve, transposed at $z$=0.9578, corresponds approximately to the observed $R-$band light curve, so in principle we should consider the $U = 365$ nm transposed light curve as the actual one observed with the $Rc = 647$ nm optical filter. These transposed light curves are shown in Fig. \ref{fig:GS}. We conclude that the  Supernova emission could have been seen between 20 and 40 days after the GRB trigger, neglecting any possible intrinsic extinction. The optical observations were made up to 7 days from the GRB trigger, reaching a limit of $R \sim 23.8$ \cite{Groot1998}, and subsequent deeper images after $\sim 60$ days \cite{Djorgovski2001}. So there are no observations in this time interval. It is appropriate to notice that the $R$-band extinction value should be large since the observed column density from the X-ray observations of the GRB afterglow is large as well \cite{Yoshida2001}: the computed light curve for the possible SN of GRB 970828 should be lowered by more than 1 magnitude, leading to a SN bump below the $R$ = 25.2 limit, see Fig. \ref{fig:GS}. The presence of very dense clouds of matter near the burst site might have darkened both the supernova emission and the GRB optical afterglow. Indeed we find the presence of clouds in our simulation at the average distances of $\sim 10^{15-16}$ cm from the GRB progenitor, with average density of $\langle n \rangle \, \approx 10^3$ part/cm$^3$ and typical dimensions of $(4 - 8) \times 10^{14}$ cm, see Fig. \ref{fig:4b.3b}.

\section{Conclusions}

In conclusion, the recent progress in the observations of X and $\gamma$-ray emission, with satellites such as Swift, Fermi, AGILE, Suzaku, Coronas-PHOTON, the possibility of observing GRB afterglows with the new generation of optical and radio telescopes, developed since 1997, and the theoretical understanding of the  BdHN paradigm, have allowed to revisit the data of GRB 970828 and give a new conceptual understanding of the underlying astrophysical scenario.

We verify in this paper that GRB 970828 is a member of the  BdHN family. This new understanding leads to a wealth of information on the different emission episodes which are observed during an IGC process.
In Episode 1, we determine the evolution of the thermal component and of the radius of the blackbody emitter, given by Eq. (1), see Figs. \ref{fig:4b.0}, \ref{fig:4b.1}. The onset of the SN is here observed for the first time in an unprecedented circunstance: a SN exploding in a close binary system with a companion NS. The energetics are correspondingly much larger than the one to be expected in an isolated SN, and presents an high energy component likely associated to an outflow process in the binary accreting system.
In Episode 2, the GRB, we give the details of the CBM structure, see Fig. \ref{fig:4b.3b}, of the simulation of the light curve and the spectrum of the real GRB emission.
We have also shown in Table \ref{tab:4b.2} the final results of the GRB simulation, the total energy of the $e^+e^-$ plasma, the baryon load $B$, the temperature of the P-GRB $kT_{th}$ and the Lorentz Gamma factor at transparency $\Gamma$, as well the average value of the CBM density $\langle n_{CBM} \rangle$ and the density ratio of the clouds $\delta n/n$.
In Episode 3, we have shown that the late afterglow emission observed by ASCA and ROSAT, although limited to few data points, when considered in the cosmological rest-frame of the emitter, presents a successful overlap with the standard luminosity behavior of other members of the BdHN family \cite{Pisani2013}, which is the most striking confirmation that in GRB 970828 an IGC process is working.
Finally, from this latter analogy with the late X-ray afterglow decay of the ``Golden Sample'' \cite{Pisani2013}, and with the optical bump observed in GRB 090618, see Fig. \ref{fig:GS}, associated to a SN emission \cite{Cano2011}, we have given reasons why a SN associated to GRB 970828 was not observable due to the large interstellar local absorption, in agreement with the large column density observed in the ASCA X-ray data \cite{Yoshida2001} and with the large value we have inferred for the CBM density distribution, $\langle n_{CBM} \rangle \approx 10^3$ particles/cm$^3$.

The possibility to observe the energy distribution from a GRB in a very wide energy range, thanks to the new dedicated space missions, has allowed to definitely confirm the presence of two separate emission episodes in GRBs associated to SNe. Future planned missions, as the proposed Wide Field Monitor detector on board the LOFT  mission \cite{LOFT2012}, will allow to observe the thermal decay from these objects down to $kT = 0.5 -1$ keV. It is important to note the possibility that the Large Area Detector, designed for the LOFT mission, will be also able to observe the afterglow emission from times larger than $10^4$ s in the rest-frame, allowing to check possible new  BdHNe by using the overlapping method described in \cite{Pisani2013,Izzo2013} and consequently estimate the distance, wherever an observed determination of the redshift is missing.

\newpage
\bibliographystyle{unsrt}

\begin{thebibliography}{10}

\bibitem{Djorgovski2001}
S.~G. {Djorgovski}, D.~A. {Frail}, S.~R. {Kulkarni}, et al.,
\newblock {\em \apj}, 562:654--663, December 2001.

\bibitem{IAUC6726}
R.~{Remillard}, A.~{Wood}, D.~{Smith}, and A.~{Levine}.
\newblock {\em \iaucirc}, 6726:1, August 1997.

\bibitem{IAUC6728}
D.~{Smith}, A.~{Levine}, R.~{Remillard}, et al., 
\newblock {\em \iaucirc}, 6728:1, August 1997.

\bibitem{IAUC6729}
A.~V. {Filippenko}, D.~{Stern}, R.~R. {Treffers}, et al.,
\newblock {\em \iaucirc}, 6729:1, August 1997.

\bibitem{IAUC6757}
J.~{Greiner}, R.~{Schwarz}, J.~{Englhauser}, et al.,
\newblock {\em \iaucirc}, 6757:1, October 1997.

\bibitem{IAUC6735}
S.~C. {Odewahn}, S.~G. {Djorgovski}, S.~R. {Kulkarni}, et al.,
\newblock {\em \iaucirc}, 6735:1, September 1997.

\bibitem{Groot1998}
P.~J. {Groot}, T.~J. {Galama}, J.~{van Paradijs}, et al.,
\newblock {\em \apjl}, 493:L27, January 1998.

\bibitem{Gehrels2004}
N.~{Gehrels}, G.~{Chincarini}, P.~{Giommi}, et al.,
\newblock {\em \apj}, 611:1005--1020, August 2004.

\bibitem{Meegan2009}
C.~{Meegan}, G.~{Lichti}, P.~N. {Bhat}, et al.,
\newblock {\em \apj}, 702:791, September 2009.

\bibitem{AGILE}
M.~{Tavani}, G.~{Barbiellini}, A.~{Argan}, et al.,
\newblock {\em \aap}, 502:995--1013, August 2009.

\bibitem{KonusWIND}
R.~L. {Aptekar}, D.~D. {Frederiks}, S.~V. {Golenetskii}, et al.,
\newblock {\em \ssr}, 71:265--272, February 1995.

\bibitem{Burrows2005}
D. Burrows, J.~Hill, J.~Nousek, et al.,
\newblock {\em Space Science Reviews}, 120:165, 2005.
\newblock 10.1007/s11214-005-5097-2.

\bibitem{Ruffini2007b}
R.~{Ruffini}, M.~G. {Bernardini}, C.~L. {Bianco},et al.,
\newblock In {\em ESA Special Publication}, volume 622 of {\em ESA Special
  Publication}, page 561, 2007.

\bibitem{Rueda2012}
J.~A. {Rueda} and R.~{Ruffini}.
\newblock {\em \apjl}, 758:L7, October 2012.

\bibitem{Izzo2012}
L.~{Izzo}, R.~{Ruffini}, A.~V. {Penacchioni}, et al.,
\newblock {\em \aap}, 543:A10, July 2012.

\bibitem{Penacchioni2012}
A.~V. {Penacchioni}, R.~{Ruffini}, L.~{Izzo}, et al.,
\newblock {\em \aap}, 538:A58, February 2012.

\bibitem[{{Hoyle} \& {Lyttleton}(1939)}]{1939PCPS...35..405H}
{Hoyle}, F., \& {Lyttleton}, R.~A. 1939, Proceedings of the Cambridge
  Philosophical Society, 35, 405.
  
  \bibitem[{{Bondi} \& {Hoyle}(1944)}]{1944MNRAS.104..273B}
{Bondi}, H., \& {Hoyle}, F. 1944, \mnras, 104, 273.

\bibitem[{{Bondi}(1952)}]{1952MNRAS.112..195B}
{Bondi}, H. 1952, \mnras, 112, 195.

\bibitem{FRR2014}
C.L. {Fryer}, J.A.~{Rueda}, R. {Ruffini}, et al.,
\newblock {\em \apjl}, 793:L36, September 2014.

\bibitem[{{Fryer} {et~al.}(1996){Fryer}, {Benz}, \&
  {Herant}}]{1996ApJ...460..801F}
{Fryer}, C.~L., {Benz}, W., \& {Herant}, M. 1996, \apj, 460, 801.

\bibitem[{{Eggleton}(1983)}]{1983ApJ...268..368E}
{Eggleton}, P.~P. 1983, \apj, 268, 368.

\bibitem[{{Fryer} {et~al.}(1999{\natexlab{a}}){Fryer}, {Benz}, {Herant}, \&
  {Colgate}}]{1999ApJ...516..892F}
{Fryer}, C.~L., {Benz}, W., {Herant}, M., \& {Colgate}, S.~A.
  1999{\natexlab{a}}, \apj, 516, 892.
  
  \bibitem[{{Chevalier}(1989)}]{1989ApJ...346..847C}
{Chevalier}, R.~A. 1989, \apj, 346, 847.

\bibitem[{{Zel'dovich} {et~al.}(1972){Zel'dovich}, {Ivanova}, \&
  {Nadezhin}}]{1972SvA....16..209Z}
{Zel'dovich}, Y.~B., {Ivanova}, L.~N., \& {Nadezhin}, D.~K. 1972, Soviet
  Astronomy, 16, 209.
  
  \bibitem[{{Houck} \& {Chevalier}(1991)}]{1991ApJ...376..234H}
{Houck}, J.~C., \& {Chevalier}, R.~A. 1991, \apj, 376, 234.

\bibitem[{{Ruffini} \& {Wilson}(1973)}]{1973PhRvL..31.1362R}
{Ruffini}, R., \& {Wilson}, J. 1973, Physical Review Letters, 31, 1362.

\bibitem[{{Fryer}(2009)}]{2009ApJ...699..409F}
{Fryer}, C.~L. 2009, \apj, 699, 409.

\bibitem[{{Fryer} {et~al.}(2006){Fryer}, {Herwig}, {Hungerford}, \&
  {Timmes}}]{2006ApJ...646L.131F}
{Fryer}, C.~L., {Herwig}, F., {Hungerford}, A., \& {Timmes}, F.~X. 2006, \apjl,
  646, L131.

\bibitem{Penacchioni2013}
A.~V. {Penacchioni}, R.~{Ruffini}, C.~L. {Bianco}, et al.,
\newblock {\em \aap}, 551:A133, March 2013.

\bibitem{Pisani2013}
G.~B. {Pisani}, L.~{Izzo}, R.~{Ruffini}, et al.,
\newblock {\em \aap}, 552:L5, April 2013.

\bibitem{Belvedere2012}
R.~{Belvedere}, D.~{Pugliese}, J.~A. {Rueda}, et al.,
\newblock {\em Nuclear Physics A}, 883:1--24, June 2012.

\bibitem{Zhang2006}
B.~{Zhang}, Y.~Z. {Fan}, J.~{Dyks}, et al.,
\newblock {\em \apj}, 642:354, May 2006.

\bibitem{Nousek2006}
J.~A. {Nousek}, C.~{Kouveliotou}, D.~{Grupe}, et al.,
\newblock {\em \apj}, 642:389, May 2006.

\bibitem{Peer2007}
A.~{Pe'er}, F.~{Ryde}, R.~A.~M.~J. {Wijers}, et al.,
\newblock {\em \apjl}, 664:L1, July 2007.

\bibitem{Meszaros2002}
P.~{M{\'e}sz{\'a}ros}.
\newblock {\em \araa}, 40:137, 2002.

\bibitem{IAUC6727}
F.~E. {Marshall}, J.~K. {Cannizzo}, and R.~H.~D. {Corbet}.
\newblock {\em \iaucirc}, 6727:1, August 1997.

\bibitem{Murakami1997}
T.~{Murakami}, Y.~{Ueda}, A.~{Yoshida}, et al.,
\newblock {\em \iaucirc}, 6732:1, September 1997.

\bibitem{Yoshida2001}
A.~{Yoshida}, M.~{Namiki}, D.~{Yonetoku}, et al.,
\newblock {\em \apjl}, 557:L27--L30, August 2001.

\bibitem{Izzo2012b}
L.~{Izzo}, J.~A. {Rueda}, and R.~{Ruffini}.
\newblock {\em \aap}, 548:L5, December 2012.

\bibitem{Piran99}
T.~Piran.
\newblock {\em Phys. Rep.}, 314:575, 1999.

\bibitem{Ruderman}
M.~{Ruderman}.
\newblock In P.~G. {Bergman}, E.~J. {Fenyves}, and L.~{Motz}, editors, {\em
  Seventh Texas Symposium on Relativistic Astrophysics}, volume 262 of {\em
  Annals of the New York Academy of Sciences}, pages 164--180, October 1975.

\bibitem{WoodsLoeb}
E.~{Woods} and A.~{Loeb}.
\newblock {\em \apj}, 453:583, November 1995.

\bibitem{LitSari2001}
Y.~{Lithwick} and R.~{Sari}.
\newblock {\em \apj}, 555:540, July 2001.

\bibitem{ZhangPeer}
B.~{Zhang} and A.~{Pe'er}.
\newblock {\em \apjl}, 700:L65--L68, August 2009.

\bibitem{Gupta}
N.~{Gupta} and B.~{Zhang}.
\newblock {\em \mnras}, 384:L11--L15, February 2008.

\bibitem{Ruffini2001L107}
R.~{Ruffini}, C.~L. {Bianco}, F.~{Fraschetti}, et al.,
\newblock {\em \apjl}, 555:L107--L111, July 2001.

\bibitem{Ruffini2001c}
R.~{Ruffini}, C.~L. {Bianco}, F.~{Fraschetti}, et al.,
\newblock {\em \apjl}, 555:L117, July 2001.

\bibitem{Christodoulou}
D.~{Christodoulou} and R.~{Ruffini}.
\newblock {\em \prd}, 4:3552, December 1971.

\bibitem{Damour}
T.~{Damour} and R.~{Ruffini}.
\newblock {\em Physical Review Letters}, 35:463, August 1975.

\bibitem{Ruffini2005b}
R.~{Ruffini}, C.~L. {Bianco}, S.-S. {Xue}, et al.,
\newblock {\em International Journal of Modern Physics D}, 14:97--105, 2005.

\bibitem{Ruffini2005}
R.~{Ruffini}, M.~G. {Bernardini}, C.~L. {Bianco}, et al.,
\newblock In {M.~Novello \& S.~E.~Perez Bergliaffa}, editor, {\em XIth
  Brazilian School of Cosmology and Gravitation}, volume 782 of {\em American
  Institute of Physics Conference Series}, page~42, August 2005.

\bibitem{Patricelli2011}
B.~{Patricelli}, M.~G. {Bernardini}, C.~L. {Bianco}, et al.,
\newblock {\em \apj}, 756:16, September 2012.

\bibitem{Cano2011}
Z.~{Cano}, D.~{Bersier}, C.~{Guidorzi}, et al.,
\newblock {\em \mnras}, 413:669--685, May 2011.

\bibitem{DellaValle2011}
M.~{Della Valle}.
\newblock {Supernovae and Gamma-Ray Bursts: A Decade of Observations}.
\newblock {\em International Journal of Modern Physics D}, 20:1745--1754, 2011.

\bibitem{Galama1998}
T.~J. {Galama}, P.~M. {Vreeswijk}, J.~{van Paradijs}, et al.,
\newblock {\em \nat}, 395:670, October 1998.

\bibitem{LOFT2012}
M.~{Feroci}, J.~W. {den Herder}, E.~{Bozzo}, et al.,
\newblock In {\em Society of Photo-Optical Instrumentation Engineers (SPIE)
  Conference Series}, volume 8443 of {\em Society of Photo-Optical
  Instrumentation Engineers (SPIE) Conference Series}, September 2012.

\bibitem{Izzo2013}
L.~{Izzo}, G.~B. {Pisani}, M.~{Muccino}, et al.,
\newblock In A.~J. {Castro-Tirado}, J.~{Gorosabel}, and I.~H. {Park}, editors,
  {\em EAS Publications Series}, volume~61 of {\em EAS Publications Series},
  pages 595--597, July 2013.

\end{thebibliography}

\end{document}